\documentclass[runningheads]{llncs}
\usepackage[T1]{fontenc}
\usepackage{graphicx}
\usepackage{comment}
\usepackage{ulem}
\usepackage{bm}
\usepackage{amsmath, amssymb}
\usepackage{type1cm}

\usepackage{url}
\usepackage{hyperref}
\usepackage{cite}

\usepackage{color}

\begin{document}
\title{Pseudo Session-Based Recommendation with Hierarchical Embedding and Session Attributes}
\titlerunning{Pseudo SBR with Hierarchical Embedding and Session Attributes}
%
\author{Yuta Sumiya\inst{1}\orcidID{0009-0002-0585-186X} \and
Ryusei Numata\inst{2}\\ \and
Satoshi Takahashi\inst{1}\orcidID{0000-0001-5573-7841}}
\authorrunning{Sumiya et al.}
%
\institute{The University of Electro-Communications, Tokyo, Japan\\
\email{\{sumiya,stakahashi\}@uec.ac.jp}\\
\and
The Japan Research Institute Limited, Tokyo, Japan\\
\email{numata.ryusei@jri.co.jp}}
\maketitle              
\begin{abstract}
Recently, electronic commerce (EC) websites have been unable to provide an identification number (user ID) for each transaction data entry because of privacy issues. Because most recommendation methods assume that all data are assigned a user ID, they cannot be applied to the data without user IDs. Recently, session-based recommendation (SBR) based on session information, which is short-term behavioral information of users, has been studied. A general SBR uses only information about the item of interest to make a recommendation (e.g., item ID for an EC site). Particularly in the case of EC sites,  the data recorded include the name of the item being purchased, the price of the item, the category hierarchy, and the gender and region of the user. In this study, we define a pseudo-session for the purchase history data of an EC site without user IDs and session IDs. Finally, we propose an SBR with a co-guided heterogeneous hypergraph and globalgraph network plus, called CoHHGN+. The results show that our CoHHGN+ can recommend items with higher performance than other methods. 
\end{abstract}
\section{Introduction}
In electronic commerce (EC) markets, the effective recommendation of items and services based on individual user preferences and interests is an important factor in improving customer satisfaction and sales, and several previous studies have focused on recommendation systems. A recommendation system is a technology that suggests items based on a user's past actions and online behavior. However, in recent years, user IDs are not assigned to users to protect their privacy. Under such circumstances, it is difficult to identify users; therefore, conventional effective recommendation systems that need user IDs cannot be used. 

Session-based recommendation (SBR), which makes recommendations without focusing on user IDs, is currently attracting attention. SBR is a method of providing recommendations based on session IDs assigned to short-term user actions. They are assigned when a user logs into an EC site, and are advantageous in that users cannot be uniquely identified as they are assigned different IDs depending on the time of day. However, even if session ID management is inadequate, there is a risk that the session ID of a logged-in user may be illegally obtained to gain access. To prevent this, we propose a new method for recommending items without using either user or session IDs. Specifically, for purchase history data to which user and session IDs are not assigned, records with consecutive user attributes, such as gender and place of residence, are defined as pseudo-sessions, and the next item to be purchased in the pseudo-session is predicted. In this manner, items that anonymous users place in their carts in chronological order can be recommended for their next purchase without using session IDs.

Generally, existing SBRs are often graph neural network (GNN)-based \cite{gnn} methods that consider only item transactions within a session. However, in the case of purchase history, other features such as item price and category tend to be observed as well. The existing method CoHHN \cite{cohhn} shows that price information and categories are effective in recommending items. In this study, we propose a new GNN model called the co-guided heterogeneous hypergraph and globalgraph network plus (CoHHGN+), which consider not only the purchase transition and price of items, but also the category hierarchy of items and auxiliary information of sessions; our model also learns the co-occurrence relationships with other sessions within the same features, and takes into account the importance of embeddings between different features and same features. In summary, our key contributions are as follows:
\begin{enumerate}
    \item A pseudo session-based high-accuracy recommendation system is proposed.
    \item We exploit session information about users and time series sales.
    \item Item hierarchies and co-occurrence relationships of the same features are considered.
\end{enumerate}

\section{Related work}
Rendle et al. proposed a Markov chain-based SBR model, called factorized personalized Markov chains (FPMC) \cite{fpmc}. FPMC is a hybrid method that combines Markov chains and matrix factorization to capture sequential patterns and long-term user preferences. The method is based on a Markov chain that focuses on two adjacent states between items and is adaptable to anonymous SBRs. However, a major problem with Markov chain-based models is that they combine past components independently, which restricts their predictive accuracy. 

Hidasi et al. proposed a recurrent neural network (RNN)-based SBR model called GRU4Rec \cite{gru4rec}. GRU4Rec models transition between items using gated recurrent units (GRUs) for inputs represented as graphs.

The purchase transitions of an EC site can be represented by a graph structure, which is a homogeneous or heterogeneous graph depending on whether the attributes of the nodes are singular or plural. Homogeneous graphs are graphs that represent relationships by only one type of node and edge and are used to represent relationships in social networks. In contrast, heterogeneous graphs are graphs that contain multiple and diverse nodes and edges and are used to represent relationships between stores and customers.

Wu et al. proposed SR-GNN \cite{srgnn}, which uses a GNN to predict the next item to be purchased in a session based on a homogeneous graph of items constructed across sessions. Using GNNs, we obtain item embeddings that are useful for predicting by introducing attention mechanism to the continuously observed item information. Currently, SBRs based on this GNN have shown more effective results than other methods, and several extended methods based on SR-GNN have been proposed. Wang et al. proposed GCE-GNN \cite{gcegnn}, which embeds not only the current session but also item transitions of other sessions in the graph.

Existing methods, such as SR-GNN and GCE-GNN are models that learn item-only transitions; however, sessions may also include item prices and categorical features. To construct a model that takes these into account, it is necessary to use heterogeneous graphs. However, when using graphs to represent the relationship between auxiliary information such as price and items, the graph becomes more complex as the number of items in a particular price range increases. Therefore, we apply an extended heterogeneous hypergraph to allow the edges to be connected to multiple nodes. This makes it possible to understand complex higher-order dependencies between nodes, especially in recommendation tasks \cite{cohhn}. Zhang et al. proposed CoHHN \cite{cohhn}, which embeds not only item transitions, but also item prices and categories. While CoHHN can consider price and item dependencies, it does not consider the hierarchical features of categories or sales information and user attributes observed during the sessions. It also does not embed the global information that represents item purchase transitions in other sessions. Therefore, we propose a new GNN model that embeds global information as in GCE-GNN, and considers item category hierarchy, user attributes, and sale information.
\section{Preliminaries}
Let $\tau$ be a feature type that changes within a given session. Let $\mathcal{V^{\tau}}=\{v_1^{\tau}, v_2^{\tau},\cdots, v_{n^{\tau}}^{\tau}\}$ be a unique set of feature $\tau$ and $n^{\tau}$ be their size. We consider four items: item ID, price, and hierarchical category of item (large and middle); we subsequently denote its item set as $\mathcal{V}^{\mathrm{id}}$, $\mathcal{V}^{\mathrm{pri}}$, $\mathcal{V}^{\mathrm{lrg}}$, and $\mathcal{V}^{\mathrm{mid}}$, respectively. Note that the prices are discretized into several price ranges according to a logistic distribution \cite{cohhn, logistic_reference}, taking into account the market price of each item.

Let $S_{a}^{\tau}=[v_1^{a, \tau}, v_2^{a, \tau}, \cdots, v_s^{a, \tau}]$ be a sequence of the feature $\tau$ for a pseudo-session and $s$ be its length. Note that each element $v_i^{a, \tau}$ of $S_{a}^{\tau}$ is belongs to $\mathcal{V}^{\tau}$. The objective of SBR is to recommend the top $k$ items from $\mathcal{V}^{\mathrm{id}}$ that are most likely to be purchased or clicked next by the user in the current session $a$.
\subsection{Heterogeneous hypergraph and global graph}
To learn the transitions of items in a pseudo-session, two different graphs are constructed from all available sessions.

We construct heterogeneous hypergraphs $\mathcal{G}^{\tau_1, \tau_2}=(\mathcal{V}^{\tau_1}, \mathcal{E}_{h}^{\tau_2})$ to consider the relationships between different features. Let $\mathcal{E}_{h}^{\tau_2}$ be a set of hyperedges for feature $\tau_2$. Each hyperedge $e_{h}^{\tau_2} \in \mathcal{E}_{h}^{\tau_2}$ can be connected to multiple nodes $v_i^{\tau_1} \in \mathcal{V}^{\tau_1}$ in the graph. This means that a node $v_i^{\tau_1}$ is connected to a hyperedge $e^{\tau_2}$ when the features $\tau_1$ and $\tau_2$ are observed in the same record. If several nodes are contained in the same hyperedge, they are considered to be adjacent.

Heterogeneous hypergraphs are a method of constructing graphs with reference to different features; however, transition regarding information about features of the same type is not considered. Additionally, item purchase transitions may include items that are not relevant to prediction. Thus, we construct the global graph shown below.

The global graph captures the relationship between items of the same type that co-occur with an item for all sessions. According to \cite{gcegnn}, the global graph is constructed based on $\varepsilon$-neighborhood set of an item for all sessions. Assuming that $a$ and $b$ are different arbitrary session, we define the $\varepsilon$-neighborhood set as follows.
\begin{equation}
    \mathcal{N}_{\varepsilon}(v_i^{a, \tau}) = \left\{v_j^{b, \tau} | v_i^{a, \tau} = {v_{i^{'}}}^{b, \tau} \in S_a^{\tau} \cap S_b^{\tau} ; v_j^{a, \tau} \in S_b^{\tau} ; j \in [i^{'} - \varepsilon, i^{'} + \varepsilon] ; a \neq b\right\},
\end{equation}
where $i^{'}$ is an index of $v_i^{a, \tau}$ in $S_b^{\tau}$ and $\varepsilon$ is a parameter that controls how close items are considered from the position of $i^{'}$ in session $B$. Consider that $\mathcal{G}_g = (\mathcal{V}^{\tau}, \mathcal{E}_g^{\tau})$ is a global graph where $\mathcal{E}_g^{\tau}$ is an edge set and $e_{g}^{\tau} \in \mathcal{E}_g^{\tau}$ connects two vertices $v_i^{\tau} \in \mathcal{V}^{\tau}$ and $v_j^{\tau} \in \mathcal{N}_{\varepsilon}(v_i^{\tau})$. Notably, the global graph only shows the relationship between identical features, and the adjacency conditions between nodes are not affected by other features.

\section{Proposed method}
\begin{figure}[t]
  \begin{center}
  \includegraphics[width=1.0\textwidth]{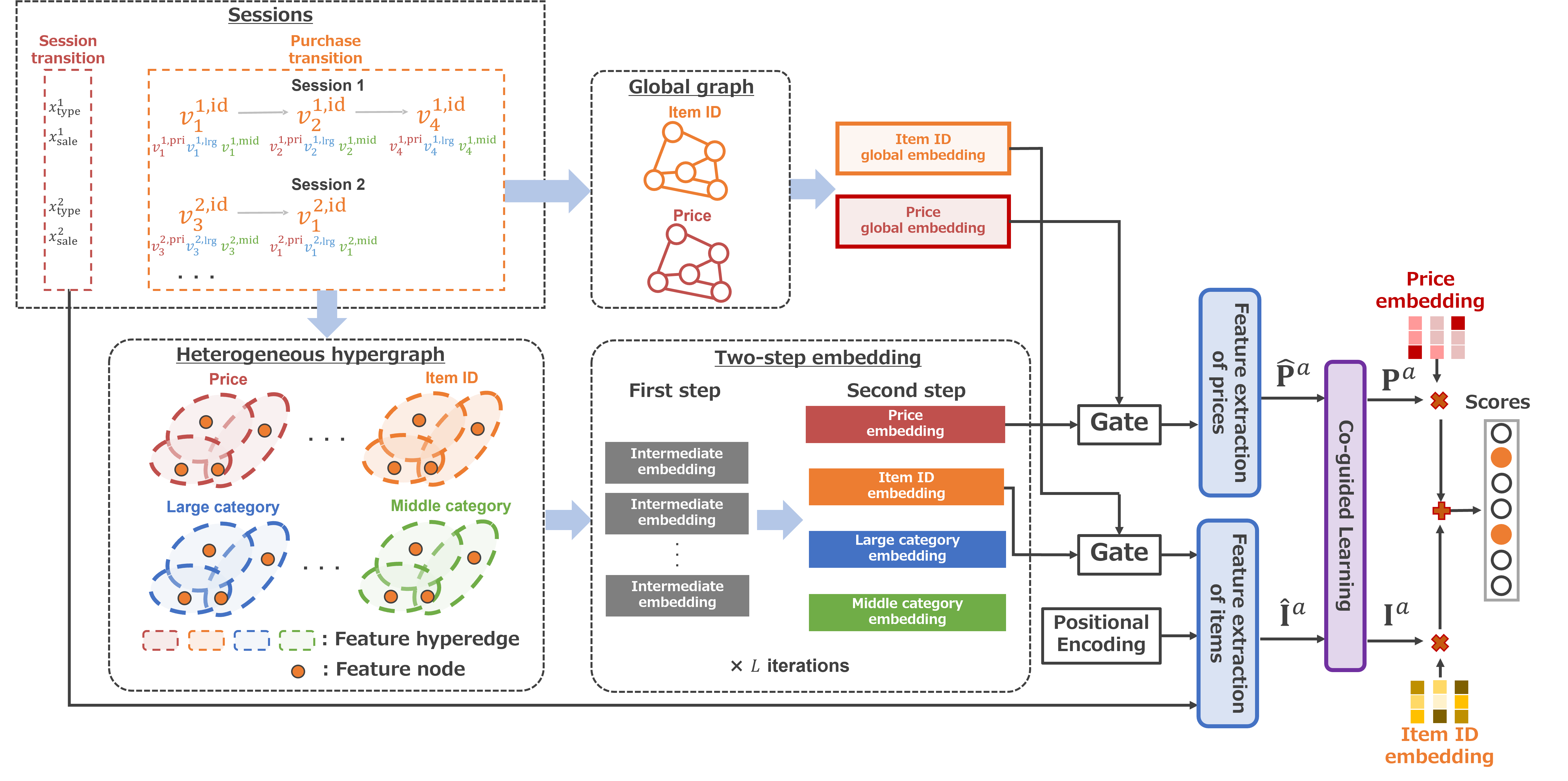}
  \caption{Overview of the proposed system. First, heterogeneous hypergraphs and global graphs are constructed for all training sessions. In two-step embedding training, embeddings within and between graphs are iteratively trained to obtain multiple feature embeddings, including categorical hierarchies. Then, using the item and price embeddings, we apply co-guided Learning \cite{cohhn} to predict the next item to be purchased by extracting features that account for transitions within the session and the interaction between the two.}
  \label{fig:causal_consistency}
  \end{center}
\end{figure}
From the perspective of privacy protection, we propose a pseudo session-based recommendation method using a heterogeneous hypergraph constructed from a set of features including a categorical hierarchy, a global graph for item and price features, and additional session attribute information. Fig. \ref{fig:causal_consistency} shows an overview of our proposed method. To consider the interactions and importance between features, our model learns feature embeddings in two steps. In the first step of aggregation, the intermediate embedding of each feature is learned from a heterogeneous hypergraph which consider the interrelationships among different features. In the second step, the final feature embedding vector obtained by aggregating the intermediate embedding in accordance with their respective importance. To address the problem of the heterogeneous hypergraph not being able to learn purchase transitions within the same feature, a global graph is used to incorporate co-occurrence relationships within the same feature into learning. Finally, we propose learning of purchase transitions within a session by considering the features of the session itself, in addition to existing methods.

\subsection{Two-step embedding with category hierarchy}
Based on intra-type and inter-type aggregating method in CoHHN \cite{cohhn}, we extend it to multiple categorical hierarchies. We obtain the item ID, price, large category, and middle category embedding vectors from the two-step learning method. In the first step of embedding, the embedding of a feature is learned from a heterogeneous hypergraph in which the feature is a node and others are hyperedges. For example, if the item ID is a node, price, large category, and middle category correspond to the hyperedges. In this case, multiple intermediate embeddings are obtained depending on the type of feature, i.e., the hyperedge. In the second step, these embeddings are used to learn the final node embeddings by aggregating them based on their importance. Each learning step is repeated for all $L$ iterations.

\subsubsection{First step}
We learn a first-step embedding for a feature $t$ from a heterogeneous hypergraph, where the target feature $t$ is a node and another feature $\tau$ is a hyperedge. First, we define the embedding of a node $v_i^t \in \mathcal{V}^{t}$ as $\bm{\mathrm{h}}_{l, i}^{\mathrm{hyper}, t} \in \mathbb{R}^d$. Here, $l$ denotes the location of the training iteration. In the initial state $l=0$, the parameters are initialized using He's method \cite{He_2015_ICCV}. Let $\mathcal{N}_\tau^{t}(v_i^t)$ be the adjacent node set of $v_i^t$. Then, the intermediate embedding of $v_i^t$ in the $l$-iteration is given by
\begin{align}
    \bm{\mathrm{m}}_{\tau, i}^{t} 
    &= \sum_{v_j^{t} \in \mathcal{N}_\tau^{t}(v_i^t)}\alpha_{j}\bm{\mathrm{h}}_{l-1, j}^{\mathrm{hyper}, t},\\
    \alpha_j &= \mathrm{Softmax}_j\left(\left[\bm{u}_{t}^{\top}\bm{\mathrm{h}}_{l-1, k}^{\mathrm{hyper}, t}\ |\ v_{k}^{t} \in \mathcal{N}_{\tau}^{t}(v_i^t)\right]\right),
\end{align}
where $\bm{u}_t^{\top}$ is an attention vector that determines the importance of $\bm{\mathrm{h}}_{l-1, j}^{\mathrm{hyper}, t}$. The function $\mathrm{Softmax}_i$ is defined as
\begin{equation}
    \mathrm{Softmax}_i\left(\left[\bm{a}_1, \cdots, \bm{a}_s\right]\right) = \frac{\exp{(\bm{a}_i)}}{\sum_{j=1}^{s}\exp{(\bm{a}_j)}}.
\end{equation}
Here, $\bm{\mathrm{m}}_{\tau, i}^{t} \in \mathbb{R}^{d}$ represents an intermediate embedding of the feature $t$ when $\tau$ is a type of hyperedge. In the first step of embedding, we learn the features to focus on when embedding $t$.

\subsubsection{Second step}
Let us assume that $\bm{\mathrm{m}}_{\tau_1, i}^{t}$, $\bm{\mathrm{m}}_{\tau_2, i}^{t}$, and $\bm{\mathrm{m}}_{\tau_3, i}^{t}$ are intermediate embeddings for a feature $t$ when $\tau_1$, $\tau_2$, $\tau_3$ are types of hyperedge, respectively. By aggregating the embeddings of the first step, we obtain the embedding of $v_i^t$ shown in the following equation.
\begin{align}
    \bm{\mathrm{h}}_{l, i}^{\mathrm{hyper}, t} 
    &= \bm{\beta}_1 * \bm{\mathrm{h}}_{l-1, i}^{\mathrm{hyper}, t} + \sum_{j=2}^{4}{\bm{\beta}_j * \bm{\mathrm{m}}_{\tau_{j-1}, i}^t},\\
    \bm{\beta}_j &= \mathrm{Softmax}_j\left(\left[\bm{W}^{t}\bm{\mathrm{h}}_{l-1, i}^{\mathrm{hyper}, t}, \bm{W}_{\tau_1}^{t}\bm{\mathrm{m}}_{\tau_1, i}^{t}, \bm{W}_{\tau_2}^{t}\bm{\mathrm{m}}_{\tau_2, i}^{t}, \bm{W}_{\tau_3}^{t}\bm{\mathrm{m}}_{\tau_3, i}^{t} \right]\right),
\end{align}
where $\bm{W}^{t}, \bm{W}_{\tau_1}^{t}, \bm{W}_{\tau_2}^{t}, \bm{W}_{\tau_3}^{t} \in \mathbb{R}^{d\times d}$ are learnable parameters, and * denotes the element-wise items of the vectors. Further, $\bm{\beta}_j$ is a parameter that computes the importance between the embedding vectors and aggregates the previous and intermediate iteration embeddings.

\subsection{Embedding of global graph}
Since heterogeneous hypergraph does not consider the co-occurrence relationships or counts between sessions related to the same feature, we use the learning of embedding global graphs in a GCE-GNN \cite{gcegnn} with two configurations: propagation and aggregation of information.
\subsubsection{Information propagation}
The $\varepsilon$-neighborhood of each feature from the global graph for feature $t$ are embedded. Because the number of features of interest within a neighborhood is considered to be different for each user, based on the attention score shown in the following equation, the neighborhood embedding $\bm{\mathrm{h}}_{\mathcal{N}_{\varepsilon}(v_i^t)}$ is first learned.
\begin{align}
    \bm{\mathrm{h}}_{\mathcal{N}_{\varepsilon}(v_i^{t})} 
     &= \sum_{v_j^{t} \in \mathcal{N}_{\varepsilon}(v_i^{t})}\pi(v_i^{t}, v_j^{t})\bm{\mathrm{h}}_{l-1, j}^{\mathrm{global}, t},\\
    \pi(v_i^{t}, v_j^{t}) &= \mathrm{Softmax}_j\left(\left[a(v_i^{t}, v_k^{t})\ |\ v_k^{t} \in \mathcal{N}_{\varepsilon}(v_i^{t})\right]\right),\\
    a(v_i^{t}, v_j^{t}) &= \bm{q}^{\top}\mathrm{LeakyRelu}\left(\bm{W}_1\left[\bm{s} * \bm{\mathrm{h}}_{l-1, j}^{\mathrm{global}, t}\right];w_{ij}\right),
\end{align}
where $\bm{\mathrm{h}}_{l-1}^{\mathrm{global}, t}$ is an embedding of the global graph for the feature $j$ on the $l-1$-th learning iteration, and $\pi(v_i^{t}, v_j^{t})$ is an attention weight that considers the importance of neighborhood node embedding. The attention score $a(v_i^{t}, v_j^{t})$ employs LeakyRelu. In LeakyRelu, $w_{ij} \in \mathbb{R}$ is the weight of an edge $(v_i^{t}, v_j^{t})$ in the global graph that represents the number of co-occurrences with features $v_j^t$, and $;$ is a concatenation operator. Further, $\bm{W}_1 \in \mathbb{R}^{(d+1)\times(d+1)}$ and $\bm{q} \in \mathbb{R}^{d+1}$ are learnable parameters, and $\bm{s}$ is the average embedding of the session to which $v_i^{t}$ belongs, defined as
\begin{equation}
    \bm{s} = \frac{1}{s}\sum_{v_i^{t} \in S_a^{t}}\bm{\mathrm{h}}_{l-1, i}^{\mathrm{global}, t}.
\end{equation}
\subsubsection{Information aggregation}
For a feature $v^{t}$ to be learned, the $l$-iteration embedding $\bm{\mathrm{h}}_{l}^{\mathrm{global}, t}$ is obtained by aggregating the $(l-1)$-iteration embedding and the neighborhood embeddings using the following formula:
\begin{align}
    \bm{\mathrm{h}}_{l, i}^{\mathrm{global}, t} 
     &= \mathrm{ReLU}\left(\bm{W}_2\left[\bm{\mathrm{h}}_{l-1, i}^{\mathrm{global}, t} ; \bm{\mathrm{h}}_{\mathcal{N}_{\varepsilon}(v_i^{t})}\right]\right),
\end{align}
where $\bm{W}_2\in\mathbb{R}^{d\times 2d}$ denotes a learnable parameter. In global graph embedding, highly relevant item information can be incorporated throughout the session by aggregating the reference features and their $\varepsilon$-neighborhoods.

\subsection{Embedding feature nodes}
For the feature node $v_{i}^{t}$, the final embedding is obtained from the embedding of heterogeneous hypergraphs considering the category hierarchy and the embedding of global graphs by the following gate mechanism:
\begin{align}
    \bm{g}_{i}^{t} &= \sigma(\bm{W}_3\bm{\mathrm{h}}_{L, i}^{\mathrm{hyper}, t} + \bm{W}_4\bm{\mathrm{h}}_{L, i}^{\mathrm{global}, t}),\\
    \bm{\mathrm{h}}_{i}^{t} &= \bm{g}_{i}^{t} * \bm{\mathrm{h}}_{L, i}^{\mathrm{hyper}, t} + (\bm{1} - \bm{g}_{i}^{t}) * \bm{\mathrm{h}}_{L, i}^{\mathrm{global}, t},
\end{align}
where $\sigma$ is a sigmoid function, $\bm{W}_3\in\mathbb{R}^{d\times d}$ and $\bm{W}_4\in\mathbb{R}^{d\times d}$ are learnable parameters, and $L$ is the final iteration of graph embedding. $\bm{g}_{i}^{t}$ is learned to consider the importance of embedding heterogeneous hypergraphs and embedding global graphs. The final feature node embedding is required only for the item ID and price based on the training of the next item.

\subsection{Feature extraction considering session attributes}
To enhance the recommendation accuracy in pseudo-sessions based on the learned node embeddings, we propose an extraction method of features related to the user's items and prices in each session.
\subsubsection{Feature extraction of items}
The embedding of an item node in session $a$ is given by the sequence $[\bm{\mathrm{h}}_{1}^{a, \mathrm{id}}, \cdots, \bm{\mathrm{h}}_{s}^{a, \mathrm{id}}]$. In addition to items, user attribute information, time-series information, and EC site sale information, among others, may be observed in each session. Therefore, we considered this information and learned to capture the session-by-session characteristics associated with the items. Let $d_\mathrm{sale}$ be the number of types of sale information and $\bm{x}_{\mathrm{sale}}^{a} \in \{0, 1\}^{d_\mathrm{sale}}$ items be given per session. Each dimension of this vector represents the type of sale, with a value of 1 if it is during a particular sale period and a value of 0 if it is outside that period. Similarly, if the number of types of attribute information is $d_\mathrm{type}$, then $\bm{x}_{\mathrm{type}}^{a} \in \{0, 1\}^{d_\mathrm{type}}$ is a vector representing user attributes.

For items and sales, we also consider time-series location information. The item location information defines a location encoding $\bm{pos\_item}_i \in \mathbb{R}^{d}$ as in \cite{attention}. Furthermore, for the location information of the sale, the week information to which the current session belongs is encoded by the following formula:
\begin{align}
    \bm{pos\_time}_{2k-1}^{a} &= \sin{\left(\frac{2m\pi}{52k}\right)},\\
    \bm{pos\_time}_{2k}^{a} &= \cos{\left(\frac{2m\pi}{52k}\right)},
\end{align}
where $\bm{pos\_time}^{a} \in \mathbb{R}^{c}$ is the location encoding associated with the week information of the session $a$, $m\in \mathbb{Z}$ represents the week, and $k$ is the embedding dimension. Because a year comprise 52 weeks, the trigonometric function argument is divided by 52. Based on the above, item embedding in a session is defined as follows:
\begin{equation}
    \bm{\mathrm{v}}_{i}^{a, \mathrm{id}} = \tanh{\left(\bm{W}_5\left[\bm{\mathrm{h}}_{i}^{a, \mathrm{id}} ; \bm{pos\_item}_i\right] + \bm{W}_6\left[\bm{x}_{\mathrm{sale}}^{a} ; \bm{pos\_time}^{a}\right] + \bm{W}_7\bm{x}_{\mathrm{type}}^a + \bm{b}_1\right)},
\end{equation}
where $\bm{W}_5 \in \mathbb{R}^{d\times2d}$, $\bm{W}_6 \in \mathbb{R}^{d\times (d_{\mathrm{sale}} + c)}$, $\bm{W}_7 \in \mathbb{R}^{d\times d_{\mathrm{type}}}$, $\bm{b}_1 \in \mathbb{R}^{d}$ are trainable parameters, $\bm{\mathrm{v}}_{i}^{a, \mathrm{id}}$ is the $i$-th item embedding in session $a$. The item preferences $\widehat{\bm{\mathrm{I}}}^{a}$ of a user in a session are determined according to \cite{cohhn} as follows:
\begin{align}
    \widehat{\bm{\mathrm{I}}}^{a} &= \sum_{i=1}^{s}\beta_i\bm{\mathrm{h}}_i^{a, \mathrm{id}},\\
    \beta_i &= \bm{u}^{\top}\sigma(\bm{W}_{8}\bm{\mathrm{v}}_{i}^{a, \mathrm{id}} + \bm{W}_{9}\bm{\mathrm{\bar{v}}}^{a, \mathrm{id}} + \bm{b}_2),
\end{align}
where $\bm{W}_{8}, \bm{W}_{9} \in \mathbb{R}^{d\times d}$, $\bm{b}_2 \in \mathbb{R}^{d}$ are learnable parameters, $\bm{u}^{\top} \in \mathbb{R}^{d}$ is the attention vector. Additionally, $\bm{\mathrm{\bar{v}}}^{a, \mathrm{id}} = \frac{1}{s}\sum_{i=1}^{s}\bm{\mathrm{v}}_{i}^{a, \mathrm{id}}$.
\subsubsection{Feature extraction of prices}
The price hyperedge in session $a$ is given by $[\bm{\mathrm{h}}_{1}^{a, \mathrm{p}}, \cdots, \bm{\mathrm{h}}_{s}^{a, \mathrm{p}}]$. To estimate price preferences with respect to users, we follow \cite{cohhn} and learn the features of the price series using multi-head attention as shown in the following equation:
\begin{align}
    \bm{\mathrm{E}}^{a, \mathrm{p}} &= [\bm{\mathrm{h}}_{1}^{a, \mathrm{p}} ; \cdots ; \bm{\mathrm{h}}_{s}^{a, \mathrm{p}}],\\
    \bm{\mathrm{M}}_{i}^{a, \mathrm{p}} &= [\bm{head}_1^{a} ; \cdots ; \bm{head}_h^{a}],\\
    \bm{head}_i^{a} &= Attention(\bm{W}_{i}^{Q}\bm{\mathrm{E}}^{a, \mathrm{p}}, \bm{W}_{i}^{K}\bm{\mathrm{E}}^{a, \mathrm{p}}, \bm{W}_{i}^{V}\bm{\mathrm{E}}^{a, \mathrm{p}}),
\end{align}
where $h$ is the number of blocks of self-attention, $\bm{W}_i^{Q}$, $\bm{W}_i^{K}$, $\bm{W}_i^{V} \in \mathbb{R}^{\frac{d}{h}\times d}$ are parameters that map item $i$ in session $a$ to query and key, value, and $\bm{head}_i^{a} \in \mathbb{R}^{\frac{d}{h}}$ is the embedding vector of each block of multi-head-attention for item $i$. Further, $\bm{\mathrm{E}}^{a, \mathrm{p}} \in \mathbb{R}^{dm}$, $\bm{\mathrm{M}}_{i}^{a, \mathrm{p}} \in \mathbb{R}^{d}$ and the embedded price series is $[\bm{\mathrm{M}}_{1}^{a, \mathrm{p}}, \cdots, \bm{\mathrm{M}}_{s}^{a, \mathrm{p}}]$.

Because the last price embedding is considered to be the most relevant to the next item price in the price series, we determine the user's price preference $\widehat{\bm{\mathrm{P}}}^{a} = \bm{\mathrm{M}}_{s}^{a, \mathrm{p}}$ in the session.

\subsection{Predicting and learning about the next item}
The user's item preferences $\widehat{\bm{\mathrm{I}}}^{a}$ and price preferences $\widehat{\bm{\mathrm{P}}}^{a}$ are transformed into $\bm{\mathrm{I}}^{a}$ and $\bm{\mathrm{P}}^{a}$ respectively by co-guided learning \cite{cohhn}, considering mutual dependency relations. When an item $v_i^{a, \mathrm{id}} \in \mathcal{V}^{\mathrm{id}}$ and a price range $v_i^{a, \mathrm{p}} \in \mathcal{V}^{\mathrm{p}}$ are observed in session $a$, the next item to view and purchase is given by the score of the following Softmax function:
\begin{align}
    \widehat{y}_i &= \mathrm{Softmax}_i\left(\left[q_1, \cdots, q_{n^{\mathrm{id}}}\right]\right),\\
    q_i &= {\bm{\mathrm{P}}^{a}}^{\top}\bm{\mathrm{h}}_{i}^{a, \mathrm{p}} + {\bm{\mathrm{I}}^{a}}^{\top}\bm{\mathrm{h}}_{i}^{a, \mathrm{id}}.
\end{align}
At the training time, this score is used to compute the cross-entropy loss.
\begin{equation}
    \mathcal{L}(\bm{y}, \widehat{\bm{y}}) = -\sum_{j=1}^{n^{\mathrm{id}}}\left(y_j\log{(\widehat{y}_j)} + (1-y_j)\log{(1-\widehat{y}_j)} \right),
\end{equation}
where $\bm{y} \in \{0, 1\}^{n^{\mathrm{id}}}$ is the objective variable that indicates whether the user has viewed and purchased item $v_i^{\mathrm{id}}$. $\widehat{\bm{y}} \in \mathbb{R}^{n\mathrm{^{\mathrm{id}}}}$ is the score for all items.
\section{Experiments}
We evaluate our proposed method using purchasing history data of an EC market. The dataset comprises the purchasing history of 100,000 people randomly selected by age group which are obtained from the users registered in 2019-20 in the Rakuten\cite{rakuten} market, which is a portal site for multiple EC sites. We consider four age groups: 21--35, 36--50, 51--65, and 66--80. Each purchasing history comprises the category name of the purchased item (large, middle, small), week (week 1--105), gender (male or female), residence (nine provinces in Japan), and price segment (separated by thousands of JPY). The user ID and session information are not recorded. Note that this dataset is provided at the 2022 Data Analysis Competition organized by Joint Association Study Group of Management Science and is not open to the public.
\subsection{Preprocessing}
\begin{table}[t]
\center
\caption{Statistical information of data set.} \label{tb:dataset_info}
\begin{tabular}{l|r|r|r|r} \hline
 Age group & \multicolumn{1}{c|}{21--35} & \multicolumn{1}{c|}{36--50} & \multicolumn{1}{c|}{51--65} & \multicolumn{1}{c}{66--80}\\ \hline
    \# of price range& 10 & 10 & 10 & 10 \\
    \# of large categories & 36 & 36 & 36 & 35 \\
    \# of middle categories & 342 & 354 & 340 & 322 \\
    \# of small categories & 2,800 & 2,975 & 2,763 & 2,327 \\
    \# of interaction & 727,655 & 1,033,405 & 712,894 & 452,496 \\
    \# of sessions & 326,110 & 462,290 & 323,184 & 203,725 \\
    Avg. session length& 2.24 & 2.24 & 2.21 & 2.24 \\ \hline
\end{tabular}
\end{table}
Our method recommends a small category name as the item ID. Additionally, the proposed model also considers session attributes, such as purchaser gender, region of residence, and EC site sales. As specific sale information, we include two types of sales that are regularly held at the Rakuten market. Sale 1 is held once every three months for one week, during which many item prices are reduced by up to half or less. Sale 2 is held for a period of one week each month, and more points are awarded for shopping for items on the EC site. Each session attribute is represented by a discrete label. When learning, we treat each gender, region, and sale as a vector with the observed value as 1 and all other values as 0. The price intervals are converted to price range labels by applying a logistic distribution \cite{logistic_reference}.

In each transformed dataset, consecutive purchase intervals with the same gender and residential area are labeled as pseudo-sessions. Based on the assigned pseudo session ID, records with a session length of less than 2 or frequency of occurrence of less than 10 are deleted, according to \cite{cohhn}. Within each session, the last observed item ID is used as the prediction target, and the other series are used for training. In dividing the data, weeks 1 through 101 are used as training data, and the remaining weeks 102 through 105 are used as test data. Additionally, 10\% of the training data re used as validation data for hyperparameter tuning of the model. The statistical details of the four datasets are listed in Table \ref{tb:dataset_info}.

\subsection{Evaluation criteria}
We employ the following criteria to evaluate the recommendation accuracy:
\begin{itemize}
    \item \textbf{P@$k$ (Precision)} : The percentage of the top $k$ recommended items that are actually purchased.
    \item \textbf{M@$k$ (Mean Reciprocal Rank)} : The mean value for the inverse of the rank of the items actually recommended for purchase. If the rank exceeds $k$, it is 0.
\end{itemize}
The precision does not consider the ranking of recommended items; however, the mean reciprocal rank is a criterion that considers ranking, implying that the higher the value, the higher the item actually purchased in the ranking. In our experiment, we set $k=10, 20$.
\subsection{Comparative model}
To verify the effectiveness of the proposed method, we compare it with the following five models.
\begin{itemize}
    \item \textbf{FPMC} \cite{fpmc} : By combining matrix factorization and Markov chains, this method can capture both time-series effects and user preferences. As the dataset is not assigned an ID to identify the user, the observations for each session are estimated as if they were separate users.
    \item \textbf{GRU4Rec} \cite{gru4rec} : An SBR based on RNN with GRU when recommending items for each session.
    \item \textbf{SR-GNN} \cite{srgnn} : An SBR that constructs a session graph and captures transitions between items using a GNN.
    \item \textbf{GCE-GNN} \cite{gcegnn} : An SBR that builds a session graph and global graph, and captures transitions between items by a GNN while considering their importance.
    \item \textbf{CoHHN} \cite{cohhn} : An SBR that constructs a heterogeneous hypergraph regarding sessions that considers information other than items and captures transitions between items with a GNN.
\end{itemize}

\subsection{Parameter setting}
To fairly evaluate the performance of the model, we use many of the same parameters for each model. For all models, the size of the embedding vector is set to 128, the number of epochs to 10, and the batch size to 100. For the optimization method, GRU4Rec uses Adagrad (learning rate 0.01) based on the results of previous studies, while the GNN method uses Adam (learning rate 0.001) with a weight decay of 0.1 applied every three epochs. The coefficients of the L2-norm regularity are set to $10^{-5}$. Additionally, in GCE-GNN and our model CoHHGN+, the size of the neighborhood item-set $\varepsilon$ in the global graph is set to 12. Furthermore, in CoHHN and our model, the number of self-attention heads is set to 4 ($h=4$), and the number of price ranges to 10. Finally, the number of GNN iterations and percentage of dropouts used in the architecture are determined by grid search for each model using the validation data. We have released the source code of our model online\footnote{https://github.com/sumugit/CoHHGN\_plus}.

\section{Results and discussion}
\begin{table}[t]
\center
\caption{
Precision of CoHHGN+ and comparative methods. The most accurate value for each dataset is shown in bold, and the second most accurate value is underlined. Each value is the average of three experiments conducted to account for variations due to random numbers. For CoHHGN+ and the other most accurate models, a t-test is performed to confirm statistical significance, and a p-value of less than 0.01 is marked with an asterisk (*).
} 
\label{tb:result_prec}
\begin{tabular}{c|cc|cc|cc|cc} \hline
Dataset & \multicolumn{2}{c|}{age 21--35} & \multicolumn{2}{c|}{age 36--50} & \multicolumn{2}{c|}{age 51--65} & \multicolumn{2}{c}{age 66--80}\\ \hline
Method & P@10 & P@20 & P@10 & P@20 & P@10 & P@20 & P@10 & P@20\\ \hline
    FPMC & 3.84 & 6.22 & 4.00 & 6.56 & 0.66 & 2.83 & 1.13 & 3.46 \\
    GRU4Rec & 1.72 & 2.57 & 1.73 & 2.60 & 1.81 & 2.82 & 1.63 & 2.73 \\
    SR-GNN & 15.06 & 20.92 & 13.71 & 20.11 & 13.78 & 20.34 & 14.30 & 22.55 \\
    GCE-GNN & 15.16 & 20.88 & 13.72 & 20.11 & 13.87 & 20.46 & \uline{14.42} & 22.46 \\
    CoHHN & \uline{15.19} & \uline{21.06} & \uline{13.96} & \uline{20.22} & \uline{13.93} & \uline{20.73} & 14.36 & \uline{22.49} \\
    CoHHGN+ & $\textbf{15.92}^{*}$ & $\textbf{22.28}^{*}$ & $\textbf{14.75}^{*}$ & $\textbf{22.01}^{*}$ & $\textbf{15.16}^{*}$ & $\textbf{22.57}^{*}$ & $\textbf{15.55}^{*}$ & $\textbf{23.84}^{*}$ \\ \hline
\end{tabular}
\end{table}
\begin{table}[t]
\center
\caption{
Mean reciprocal rank of CoHHGN+ and comparative methods. The symbols attached to the values are the same as those in the table \ref{tb:result_prec}.
} 
\label{tb:result_mrr}
\begin{tabular}{c|cc|cc|cc|cc} \hline
Dataset & \multicolumn{2}{c|}{age 21--35} & \multicolumn{2}{c|}{age 36--50} & \multicolumn{2}{c|}{age 51--65} & \multicolumn{2}{c}{age 66--80}\\ \hline
Method & M@10 & M@20 & M@10 & M@20 & M@10 & M@20 & M@10 & M@20\\ \hline
    FPMC & 0.88 & 1.04 & 1.14 & 1.31 & 0.15 & 0.28 & 0.38 & 0.54 \\
    GRU4Rec & 0.78 & 0.84 & 0.75 & 0.81 & 0.71 & 0.78 & 0.59 & 0.66 \\
    SR-GNN & 6.56 & 6.95 & 5.95 & 6.38 & 5.51 & 5.97 & 5.24 & 5.80 \\
    GCE-GNN & 6.65 & 7.04 & \textbf{6.02} & \textbf{6.45} & 5.58 & 6.04 & 5.21 & 5.75 \\
    CoHHN & \uline{6.67} & \uline{7.06} & \uline{6.01} & \uline{6.44} & \uline{5.62} & \uline{6.08} & \uline{5.27} & \uline{5.81} \\
    CoHHGN+ & $\textbf{6.89}^{*}$ & $\textbf{7.32}^{*}$ & 5.93 & 6.42 & $\textbf{5.83}^{*}$ & $\textbf{6.34}^{*}$ & $\textbf{5.81}^{*}$ & $\textbf{6.37}^{*}$ \\ \hline
\end{tabular}
\end{table}

\subsection{Performance comparison}
Tables \ref{tb:result_prec} and \ref{tb:result_mrr} show the results of evaluating the five existing methods and the proposed method CoHHGN+ on the four selected datasets. CoHHGN+ obtains the most accurate results for all datasets with precision for $k=10, 20$. The mean reciprocal rank is also the most accurate, except for the data for the 36--50 age group. For the 36--50 year age group dataset, the precision is higher than that for the other models, while the mean reciprocal rank shows the highest accuracy for GCE-GNN. However, there is no statistically significant difference in the prediction accuracy between CoHHGN+ and GCE-GNN in this dataset. Thus, it can be inferred that there is no clear difference in prediction accuracy. This confirms the effectiveness of the proposed method for all the data.

In the comparison method, a large discrepancy in accuracy between the GNN-based method, which introduces an attention mechanism in the purchase series, and the other methods is noted. Overall, the GRU4Rec without attention mechanism results in the lowest accuracy, suggesting that the results were not sufficiently accurate for data with a small number of sessions. This is because the model focuses only on purchase transitions between adjacent items. Similarly, for FPMC, although the accuracy is improved compared to GRU4Rec, modeling with Markov chains and matrix factorization is not effective for purchase data with pseudo-sessions. Moreover, SR-GNN, GCE-GNN, CoHHN, and CoHHGN+ using graphs of purchase transitions between sessions show a significant improvement in accuracy and are able to learn the purchase trends of non-adjacent items as well.

Among the compared methods, CoHHN, which considered price and large category information in addition to item ID information, tends to have a higher prediction accuracy overall. The number of series per session is generally small for purchase history data, and it can be said that higher accuracy can be obtained by learning data involving multiple features, including items. GCE-GNN, which also considers the features of other sessions, shows the second highest prediction accuracy after CoHHN. When using purchase history data with short session lengths, it is more accurate to learn embedding vectors by considering items that have co-occurrence relationships with other sessions, in addition to series within sessions. The SR-GNN that has learned only from item ID transitions is inferior to the GCE-GNN in terms of overall accuracy among GNN-based systems, although it is more accurate than the GCE-GNN for some datasets. Therefore, it can be considered that adopting features other than the item ID and other session information will lead to improved recommendation accuracy.

We confirme that the proposed method improves accuracy not only by considering auxiliary information in the purchase transition of items, but also by learning methods for its embedding vectors and including additional features that change from session to session. Furthermore, the embedding vector obtained from the global graph of the item of interest works well for a series with short session lengths.

\subsection{Impact of each model extension}
Next, we conduct additional experiments on four datasets to evaluate the effectiveness of embedding item category hierarchies and accounting for session attributes, as well as global-level features. Particularly, we design the following two comparative models:
\begin{itemize}
    \item CoHHGN (H): A model that incorporates hierarchical embedding of three or more features that vary within a session.
    \item CoHHGN (HS): A model that considers the hierarchical embedding of three or more features and session attributes in the proposed method.
\end{itemize}
To compare the performance with existing methods, we use the most accurate values of the existing methods shown in Tables \ref{tb:result_prec} and \ref{tb:result_mrr} as the baselines. Tables \ref{tb:proposal_prec} and \ref{tb:proposal_mrr} show the prediction results of the comparison model. For both precision and Mean Reciprocal Rank, CoHHGN+, which incorporates all the proposed methods, performs better overall than the other two models. For Precision, the accuracy of CoHHGN (HS) is higher for P@10 in the 21--35 year age group dataset. However, because the accuracy of CoHHGN+ is higher than that of other methods in P@20, we believe that considering the embedding of global graph features will improve the accuracy in a stable manner. For CoHHGN (H), although the accuracy is improved over the baseline in several datasets, no statistically significant differences are identified. However, extending the model to CoHHGN (HS), which also considers session attributes, results in a significant difference in precision in all datasets, except for the age group 51--65.

Further, considering the mean reciprocal rank, although the recommendation accuracy tends to improve as the model is extended to CoHHGN (H) and CoHHGN (HS), the only dataset in which statistically significant differences can be confirmed is that for the 66--80 age group. However, when extended to CoHHGN+, which incorporates all the proposed methods, the overall prediction accuracy is higher and significant differences are confirmed. This confirms that the recommendation accuracy of the item ID can be improved by simultaneously considering features that vary between sessions and attributes of other sessions, in addition to features that vary within sessions.

\section{Conclusion}
\begin{table}[t]
\center
\caption{
Comparison of the precision accuracy for each model extension. The most accurate values for each dataset are shown in bold. Each value is the average of three experiments conducted to account for random number variation. A t-test was conducted to confirm the statistical significance of the accuracy between the baseline and the proposed method, and an astarisk (*) is added if the p-value is less than 0.01.
} \label{tb:proposal_prec}
\begin{tabular}{c|cc|cc|cc|cc} \hline
Dataset & \multicolumn{2}{c|}{age 21--35} & \multicolumn{2}{c|}{age 36--50} & \multicolumn{2}{c|}{age 51--65} & \multicolumn{2}{c}{age 66--80}\\ \hline
Method & P@10 & P@20 & P@10 & P@20 & P@10 & P@20 & P@10 & P@20\\ \hline
    Baseline & 15.19 & 21.06 & 13.96 & 20.22 & 13.93 & 20.73 & 14.42 & 22.49\\
    CoHHGN (H) & 15.24 & 21.13 & 14.10 & 21.13 & 13.98 & 20.69 & 14.24 & 22.56  \\
    CoHHGN (HS) & $\textbf{15.95}^{*}$ & $22.11^{*}$ & $14.66^{*}$ & $21.97^{*}$ & 13.97 & 20.57 & $15.13^{*}$ & $23.51^{*}$  \\
     CoHHGN+ & $15.92^{*}$ & $\textbf{22.28}^{*}$ & $\textbf{14.75}^{*}$ & $\textbf{22.01}^{*}$ & $\textbf{15.16}^{*}$ & $\textbf{22.57}^{*}$ & $\textbf{15.55}^{*}$ & $\textbf{23.84}^{*}$ \\ \hline
\end{tabular}
\end{table}
\begin{table}[t]
\center
\caption{
Comparison of mean reciprocal rank accuracy for each model extension. The symbols attached to the values are the same as those in the table \ref{tb:proposal_prec}.
} \label{tb:proposal_mrr}
\begin{tabular}{c|cc|cc|cc|cc} \hline
Dataset & \multicolumn{2}{c|}{age 21--35} & \multicolumn{2}{c|}{age 36--50} & \multicolumn{2}{c|}{age 51--65} & \multicolumn{2}{c}{age 66--80}\\ \hline
Method & M@10 & M@20 & M@10 & M@20 & M@10 & M@20 & M@10 & M@20\\ \hline
    Baseline & 6.67 & 7.06 & 6.02 & 6.45 & 5.62 & 6.08 & 5.27 & 5.81\\
    CoHHGN (H) & 6.63 & 7.02 & \textbf{6.02} & 6.45 & 5.66 & 6.12 & 5.32 & 5.88  \\
    CoHHGN (HS) & 6.77 & 7.19 & 5.94 & \textbf{6.45} & 5.65 & 6.11 & $5.69^{*}$ & $6.24^{*}$  \\
    CoHHGN+ & $\textbf{6.89}^{*}$ & $\textbf{7.32}^{*}$ & 5.93 & 6.42 & $\textbf{5.83}^{*}$ & $\textbf{6.34}^{*}$ & $\textbf{5.81}^{*}$ & $\textbf{6.37}^{*}$ \\ \hline
\end{tabular}
\end{table}
In this study, we developed CoHHGN+ based on CoHHN, which is an SBR considering various features, and GCE-GNN considering global graphs, for purchase history data of EC sites. Moreover, we considered global time-series information, sale information, and user information. The application of the proposed model to pseudo-session data with no user IDs shows that the GNN-based method exhibits significantly higher accuracy than those for the other methods, and that our proposed CoHHGN+ is the most accurate method on the dataset.

Althought incorporating several types of data improves the prediction accuracy, there are still issues from the viewpoint of feature selection for data with more types of information recorded. If there are $n$ types of heterogeneous information, the number of heterogeneous hypergraphs used to embed heterogeneous information is $2^n$. Therefore, selecting and integrating heterogeneous information remains an issue.

Future work on issues related to more efficient feature selection and methods for integrating heterogeneous information will lead to the development of models with even higher accuracy. We would also like to expand the scope of application of CoHHGN+ proposed in this study and attempt to provide useful recommendations in other domains as well.

\subsubsection*{ACKNOWLEDGMENTS} 
We would like to thank the sponsor of the Data Analysis Competition, Joint Association Study Group of Management sCience (JASMAC), and Rakuten Group, Inc. for providing us with the data. This work was also supported by JSPS KAKENHI Grant Number JP20H04146.

\bibliography{reference} 
\bibliographystyle{splncs04}

\end{document}